\documentclass[12pt]{article}

\usepackage{amssymb,latexsym,amsmath} 
\usepackage{authblk}
\usepackage[utf8]{inputenc}
\usepackage{dsfont}
\usepackage[utf8]{inputenc}
\usepackage{graphics}
\usepackage{color}
\usepackage{subfig}
\usepackage{graphicx}
\usepackage[colorlinks = true,
linkcolor = blue,
urlcolor  = blue,
citecolor = blue,
anchorcolor = blue]{hyperref}

\begin{document}
\title{On Maximum Complexity in Holography}

\author{Shahrokh Parvizi \thanks{parvizi@modares.ac.ir} }
\author{Mojtaba Shahbazi \thanks{mojtaba.shahbazi@modares.ac.ir} }

\affil{Department of Physics, School of Sciences,
	Tarbiat Modares University, P.O.Box 14155-4838, Tehran, Iran}

\maketitle
\begin{abstract} 
In a quantum circuit, it is believed that complexity itself reaches a maximum of order exponential in the number of q-bits or equivalently exponential in entropy of the black hole. However, the current holographic proposals do not meet this criterion. The holographic proposals find the complexity of the very late times to be linear in the entropy, while in the quantum circuit, it is expected that complexity meets within a finite time its maximum value in an exponential in the entropy. These points are required to be altered in holographic proposals of complexity. This paper introduces a new holographic proposal that meets this criterion and consolidates the Lloyd bound.

\end{abstract}

\section{Introduction}
There are some paradoxes in the combination of quantum mechanics and the general theory of relativity, mainly when one focuses on the physics of black holes by quantum mechanics' eye \cite{amps}. In this manner, the study of black hole physics may provide some meteoric way to a successful quantum gravity theory. Recently Susskind's proposal based on the AdS/CFT correspondence made a big jump considering the interior of black holes \cite{shock}. By this conjecture, the volume of Einstein-Rosen bridges (ERB) corresponds to the complexity of the conformal field theory on the boundary of an AdS space, which is coined as Complexity=Volume (CV) proposal. Complexity is the minimum number of quantum gates required to turn a reference state into a target state.\\ Later, Brown et al. \cite{action1,action2} suggested another holographic proposal dubbed as Complexity=Action (CA) proposal. It states that the action of bulk theory in the Wheeler-DeWitt patch, which is the domain of all space-like trajectories anchored to either side of the Penrose diagram is dual to the complexity of the boundary theory.

It can be shown that in any quantum circuit, there is an upper bound of $e^K$ for the complexity, where $K$ is the number of q-bits \cite{shock}. Because $K$ number of q-bits is proportional to the entropy $s$, we put $K=s$. In this way, the maximum complexity is $C_{max}=e^s$ \cite{shock,three lecture}. Furthermore, the complexity growth rate is bounded from above by the mass of the system known as the Lloyd bound \cite{lloyd}.

It has been revealed that some theories do not respect the Lloyd bound by CV and some by CA proposal in the full-time behavior \cite{onthe}, in Einstein-Maxwell-Dilaton theories \cite{ca2,hbound}, in Lifshitz hyper scaling violating exponent space \cite{ali}, in local quantum quench \cite{local} and an anisotropic solution of Einstein-Dilaton-Axion theory \cite{aniso}. This point leads to new proposals \cite{ca2,couch} and modifications \cite{ref,al,vol}.

Nonetheless, none of these cases could maintain the criterion of maximum complexity because they find the complexity for the late times linear in entropy. Indeed, maximum complexity arises at times of order exponential in entropy. Besides, the bulk side in AdS/CFT can not afford the quantum effects at late times regime \cite{eter}. The point is that by Poincaré recurrence theorem, every unitary evolution by finite entropy will return to its initial state at late but finite times\footnote{Non-unitary evolutions with finite entropy are not required to respect Poincaré recurrence but it could be managed to have the recurrences by nonperturbative dynamical effects \cite{barbon}.}. On the bulk side, correlation functions at late times are to be returned to a lower bound to respect the Poincaré recurrence; however, one can not recover Poincaré recurrence in the presence of black holes \cite{eter}. On its face, the maximum complexity that is reached at late times (the classical Poincaré recurrence time) is a quantum effect and can not be recovered on the bulk side. Nonetheless, Maldacena's proposal in \cite{eter} by some averaging pursues recovering the recurrences at classical Poincare recurrence time; in addition, in two-dimensional gravitational theory, it could be shown that there would be the maximum complexity by considering the full path integral. Then it makes sense to consider maximum complexity in holography.

On the other hand, due to the ambiguity in the gate choice in the holographic complexity, there would be a class of holographic proposals that all members share the same expected requirements, then there is not a unique holographic proposal of the complexity. In this paper, we provide a new class of holographic proposal given in \eqref{hc}, that can fulfill the maximum complexity condition, and we will see that it behaves the Lloyd condition well, particularly for the large masses. 

Our proposal, which we call ``Hyperbolic Proposal" could be seen either as an averaged complexity. In Sachdev-Ye-Kitaev (SYK) models that are examples of chaotic systems, the hyperbolic proposal bears a resemblance to averaged complexity.

The organization of the paper is as follows: In the next section, the expected features of holographic complexity candidates are reviewed. Our new proposal is introduced in section \ref{sec:HP}, and we discuss its behavior. In section \ref{syk}, the relation between the hyperbolic proposal and averaged complexity is drawn, and we conclude in the final section.
\section{Review on Expected Features of Holographic Complexity}
It is known that the computational complexity represents four expected features: linear growth, a particular behavior under perturbation, a maximum value of complexity, and the Lloyd bound \cite{three lecture,expo}. As a consequence, it is expected that the holographic picture of complexity embraces these features.

One may consider the holographic complexity as a function like $C(\lambda)$, where $\lambda$ is a dynamical variable in the black hole and $C$ is a suitable function. In the case of well-known CV and CA proposals, $\lambda$ is respectively the volume of the Einstein-Rosen bridge and the action in the Wheeler-DeWitt patch, and $C$ is simply a linear function with constant coefficients. 

First of all, we expect the holographic complexity $C(\lambda)$ to show a linear growth in time and, secondly, behaves under perturbation as its dual in quantum circuits. Although it is expected that complexity reaches its maximum in a finite time, the current holographic proposals lack this condition. 

Let us discuss these behaviors in the following subsections and review the expected features in CV and CA proposals.
\subsection{Linear growth}\label{lineargr}
In quantum circuits, it can be seen that restricted to a shorter time-scale when complexity is much smaller than its maximum, its growth is linear \cite{shock}:
$$C\propto Kt$$
where $K$ is the number of q-bits. Although except at the beginning, it could be non-linear, it is believed that complexity grows all the way to its maximum linearly \cite{expo}. 

Let us start with the CV proposal and consider a black hole geometry of the following form:
\begin{align}
ds^2&=-f(r) dt^2+\frac{dr^2}{f(r)}+r^2d\Omega_{D-2}^2 \\
f(r)&=1+r^2-\frac{\mu}{r^{D-3}}
\end{align}
where for $D>3$,
\begin{align}
\mu=\frac{16\pi G_N M}{(D-2)\omega_{D-2}}
\end{align}
in which $\omega_{D-2}$ is the volume of $D-2$ dimensional sphere. The volume growth rate of ERB which is defined for a two-sided black hole is given by \cite{shock}:
$$\frac{dV}{dt}=\omega_{D-2}r^{D-2}\sqrt{f(r)}$$
By induced metric for a volume in AdS-Schwarzschild, the volume is written as \cite{shock}:
\begin{equation}
V=\int r^2\sqrt{-f+\frac{r'^2}{f}} dtd\Omega \label{v}
\end{equation}
where $r'=\frac{dr}{dt}$. It could be shown that for finite times Eq. (\ref{v}) turns into \cite{shock}:
\begin{equation} \label{vol-linear}
V\propto sT|t_L+t_R|
\end{equation}
where $s$ and $T$ are entropy and temperature of the black hole, respectively, and $t_L$ and $t_R$ are times for the left and right side of the black hole. With $C\propto V$, Eq. \eqref{vol-linear} indicates linear growth of the holographic complexity.

In the CA proposal, the computational complexity is proportional to the action of the theory in the WDW patch, $C\propto \mathcal{S}$. Since the action is computed in a bounded manifold, the Gibbons-Hawking-York term should be added to the action of the theory:

$$\mathcal{S}=\mathcal{S}_{theory}+\frac{1}{8\pi G}\int_{\partial M}\sqrt{h}K$$
where $h$ is the induced metric on the boundary of the patch and $K$ is extrinsic curvature constructed by the induced metric. At late but finite times, the complexity growth rate for a $D$ dimensional AdS-Schwarzschild is given by \cite{action2}:
\begin{equation}\label{dSdt}
\frac{d\mathcal{S}}{dt}=-\frac{\Omega_{D-2}r_h^{D-1}}{8\pi G L_{AdS}^2}=2sT
\end{equation}
This shows the linear behavior as expected.

The expressions \eqref{vol-linear} and \eqref{dSdt} show that the current proposals of holographic complexity describe the linear growth of complexity.

\subsection{Perturbation}
\begin{figure}
\centering
\includegraphics[width=10cm]{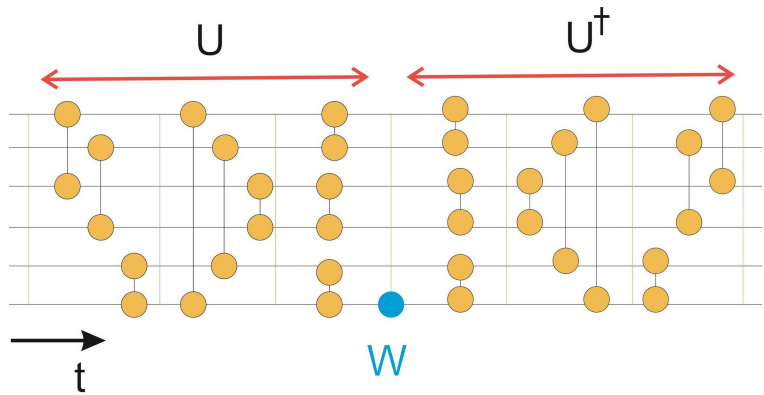}
\caption{A circuit of precursor. The horizontal lines are q-bits and the circles are quantum gates.}\label{pre}
\end{figure}

In \cite{epr}, Susskind and Maldacena showed that by holography a pair of entangled particles is dual to an ERB or in short $ER=EPR$. Then a perturbation in gravity leads to a perturbation in the pair particles. Another test to check is on the perturbed thermo-field double (TFD) states which are dual of the geometries with shock waves \cite{shock,multi}. In other words, a perturbation in the evolution of the black hole should hold in the correspondence between gravity and the quantum circuit. Small perturbations in TFD state could be conducted by precursors as:
$$|\psi(t)\rangle=W(t)|\psi\rangle$$
where $W(t)=U^{\dagger}WU$ with $W$ close to one and $U=e^{-iH}$. The schema of a quantum circuit of the precursors could be seen in Fig. \ref{pre}. 

\begin{figure}
\centering
\includegraphics[width=7cm]{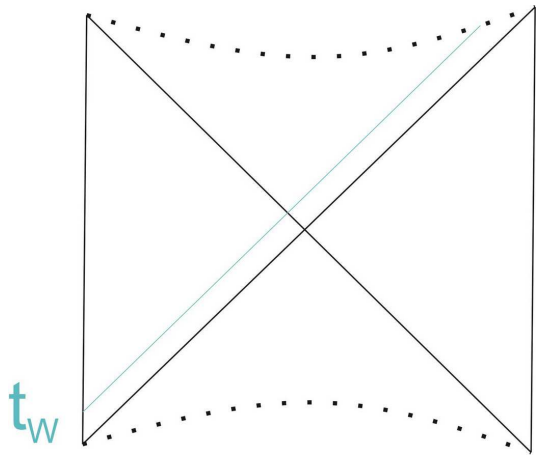}
\caption{A Penrose diagram for a two-sided black hole with a shock wave at time $t_w$. This is supposed to be dual to a precursor.}\label{sho}
\end{figure}

The complexity of the perturbed TFD state is read as \cite{shock}:
\begin{align}\label{C1-perturb}
C\propto K(t_f-2n_pt_*)
\end{align}
where $t_f$ is the total time of the perturbation, $n_p$ is the number of stages of the perturbation and $t_*$ is the scrambling time. The dual picture is the geometries with $n_p$ number of shock waves as in Fig. \ref{sho} \cite{shock}. 

By CV proposal the volume of ERB in the presence of $n_p$ number of shock waves is given by \cite{shock}:
\begin{align}\label{C2-perturb}
V\propto (t_f-2n_pt_*)+ O(1)
\end{align}
where $O(1)$ is the various time variables terms of the first order.

In CA proposal, where the action of the theory in the presence of $n_p$ number of shock waves should be considered, the action at late times is read as \cite{action2}:
\begin{align}\label{C3-perturb}
\mathcal{S}\propto (t_f-2n_pt_*)
\end{align}
At finite times there would be some corrections but the leading term is linear. Eqs. \eqref{C2-perturb} and \eqref{C3-perturb} indicate that both CV and CA proposals are in agreement with the perturbed TFD in \eqref{C1-perturb}.

\subsection{Lloyd bound}
The complexity growth rate is bounded from above by a conjecture called the Lloyd bound \cite{lloyd}:
\begin{equation}
\frac{dC}{dt}\le\frac{2M}{\pi} \label{lloyd}
\end{equation}
where $M$ is the mass of the black hole. This inequality originates in the uncertainty principle on the quantum side. In other words, the time a quantum gate takes to operate on a state is restricted by the uncertainty in energy and time. The holographic models should respect the inequality. However, there are some theories that either CV or CA violates the bound \cite{onthe}-\cite{aniso}. Although there is a new proposal called CV-2.0 \cite{couch}, which can improve the Lloyd bound condition, particularly for Einstein-Maxwell Dilaton theories \cite{ca2}, it also suffers from some violating cases \cite{cv2v}. In section \ref{sec:HP}, we show how it can be improved.

\subsection{Maximum complexity}
In a quantum circuit, computational complexity reaches its maximum of exponential in entropy at a finite time and fluctuates around its maximum. Furthermore, after quantum Poincaré recurrence time, it jumps off to its initial value as Fig. \ref{qc} depicts \cite{three lecture}. It is expected that a holographic picture of complexity would exhibit the maximum value. However, it is believed that the general theory of relativity breaks down at the classical Poincaré recurrence time\footnote{The classical Poincaré time is of order exponential in entropy while quantum Poincaré recurrence time is of order double exponential in entropy \cite{three lecture}.} where complexity reaches its maximum. Since the maximum value of complexity occurs at the classical Poincaré recurrence time, none of the holographic proposals of complexity shows the
maximum value i.e. exponential in entropy.\\

There are two related reasons that general relativity breaks down at the classical Poincaré recurrence time. The first reason is due to the correlation functions of opposite boundaries in extended AdS black holes at the classical Poincaré recurrence time. It seems that the correlation functions at those times should reach a minimum of order $e^{-\alpha s}$ where $\alpha$ is a constant and $s$ the entropy of the black hole; however, on the bulk side, it lacks the minimum and vanishes. Nonetheless, one can recover the minimum at the classical Poincaré recurrence time by non-perturbative dynamical effects\footnote{In fact, the path integral on the bulk side is over some geometries with the same boundary condition. It could be seen that there are geometries that recover the minimum and have small free energy and large contribution compared to the geometries that do not recover the minimum.} and some averaging \cite{eter,barbon}\footnote{For a breakdown of the general theory of relativity at the quantum recurrence time see \cite{stan}}. The second reason stems from the finite value of entropy of a black hole. The correlation functions of opposite boundaries are given by $\langle\mathcal{O}_L\mathcal{O}_R\rangle\sim e^{-\gamma t}$ where $\gamma$ is a constant \cite{three lecture, eter}. It seems that the correlations are related to ERBs and as time elapses, the length of ERB grows, which makes the correlations decay. Since the entropy is finite, as time marches on, the evolution of quantum states passes through a finite series of orthogonal states. Consequently, the time it takes to run out of orthogonal states is of order exponential in entropy and after that, because the states become superposition of previous states, then the correlations stop decreasing but on the bulk side $e^{-\gamma t}$ continues to zero \cite{three lecture}. 

While the general theory of relativity breaks down where complexity reaches its maximum, any attempt to recover maximum complexity in holography could be seen as an effective description of the quantum phenomenon in a semi-classical context. In principle, it can be understood as considering the quantum effects in a path integral approach which yields to an effective action of classical fields. The details of quantum effects and their calculation are not known, but there are criteria and evidences which may guide us to the correct form of effective classical terms. The same procedure takes place where entanglement entropy of the Hawking radiation is considered. There is a lot in common between entanglement entropy and complexity when their developments are viewed. On one hand, the celebrated Ryu-Takayanagi (RT) surface links entanglement entropy (EE) of the boundary quantum field theory with the bulk surface, on the other hand, holographic complexity links the complexity of operators in the quantum field theory with the volume of the ERB (in CV proposal). It has been seen that RT surface does not account for the loop correction \cite{malda}, in other words, there is a quantum correction that the holographic EE cannot recover. As a counterpart, the holographic complexity cannot recover the maximum complexity which is a quantum effect. In addition, there is a prescription in the low energy limit of quantum gravity that recovers the loop correction of EE known as quantum extremal surface (QES) \cite{penin,alm}. As a counterpart, in the next section, we provide a prescription that recovers the maximum complexity in the low energy limit of quantum gravity, we call it Hyperbolic proposal. And finally, the full path integral computation in the low dimensional gravity shows that the QES naturally arises \cite{sta,maldarep}. As a counterpart, the full path integral computation in the low dimensional gravity shows that the maximum complexity arises \cite{mcom}.

\begin{center}
\begin{tabular}{ |c|c| } 
 \hline
\textbf{Entanglement Entropy} & \textbf{Complexity} \\ 
 RT Surface & CV/CA/CV-2.0 \\
Loop Correction & Maximum Complexity \\
 QES & Hyperbolic Proposal  \\
Full Path Integral& Full Path Integral\\
 \hline
\end{tabular}
\end{center}

\section{Hyperbolic Proposal}\label{sec:HP}
As we discussed in the last section, the CV and CA proposals and their extensions do not properly fit all criteria for a holographic complexity. Here we make a modification which can improve the situation. Holographic proposals for complexity are a function of a dynamical variable $\lambda$, $\mathcal{C}(\lambda)$ where the dynamical variable can be either volume $V$ or action $A$ as in the CV and CA proposals or any modification of them. In this new proposal, we consider $\lambda=C$ where $C$ is either previous proposal. The main point is adopting $\mathcal{C}$ function to be a hyperbolic function instead of a linear one. 

It should be noted that there is an ambiguity in the notion of holographic complexity which corresponds to the ambiguity in the choice of the gate set in the complexity theory and there would be a class of holographic proposals that share the linearity and the particular behavior under the perturbation discussed in the previous section, put it differently, the time dependence in holographic complexity is important \cite{ambi}. Therefore, there is no uniqueness in the holographic complexity. In \cite{ambi}, a class of CV proposal is introduced as:
\begin{equation}\label{class}
C\propto \int_{\Sigma_{ F_2}} d^dx \sqrt{g_{ind.}} F_1(g_{\mu\nu},\sigma^{\mu})
\end{equation}
where $F_1$ is a scalar function of the induced metric $g_{ind.}$ and an embedding $\sigma^{\mu}$ of the codimension-one surface $\Sigma _{F_2}$, which in turns comes from extremizing a similar functional replacing $F_1$ with another scalar function $F_2$. As long as the criteria of the holographic complexity are satisfied, functions $F_1$ and $F_2$ are arbitrary. 
When $F_1=F_2=1$, \eqref{class} reduces to the known CV proposal.

Now we are claiming such an arbitrariness not inside the functional \eqref{class} but outside of it, i. e., $\mathcal{C}(C)=\mathcal{C}(\int_{\Sigma_{ F_2}} d^dx \sqrt{g_{ind.}} F_1(g_{\mu\nu},\sigma^{\mu}))$. Here, we may take $F_1=F_2=1$ and look for a proper function $\mathcal{C}$ to satisfy all criteria mentioned in the previous section.

This is inspired by studying the complexity in a BTZ black hole which can be computed analytically. This makes it a fertile ground to see the explicit behavior of the complexity, which is given by \cite{onthe}:
\begin{equation}
C=s \bigg(1+log\Big(\frac{\omega}{T}\cosh(\omega T t)\Big)\bigg) \tanh(\omega T t)\label{btz}
\end{equation}
where $\omega$ is a constant. At early times, the time dependence of \eqref{btz} is $C\propto \tanh(\omega T t) $
which is linear around $t\simeq 0$ at leading order. On the other hand, the complexity in the linear region behaves as $C\sim sTt$. So inspired by the BTZ black holes, we can consider a new proposal with introducing function $\mathcal{C}$ to be a hyperbolic function of $\omega C$. It would be a new class of holographic complexity which we call the class of {\it 'Hyperbolic Proposals'}, and introduce it as:
\begin{equation}\label{hc}
\mathcal{C}=A \frac{\sinh(\omega_1 C-\alpha)}{\cosh(\omega_2 C-\beta)}
\end{equation}
where $C$ is either CV or CA proposal. If $\omega_1=\omega_2=\omega$ and $\alpha=\beta=0$ it yields to the complexity in the BTZ black hole \eqref{btz}. This function would be quite inspiring when we consider Sachdev-Ye-Kitaev (SYK) model in section \ref{syk}. 

As an example we are going to consider a member of this class with parameters $A=1$, $\omega_1=\omega_2=\text{sech}(s)=:\omega$,  $\alpha=0$ and $\beta=s$, the entropy. Then
\begin{equation} \label{hp1}
\mathcal{C}=\frac{\sinh(\omega C)}{\cosh(\omega C-s)}
\end{equation}

\begin{figure}
\centering
\includegraphics[width=10cm]{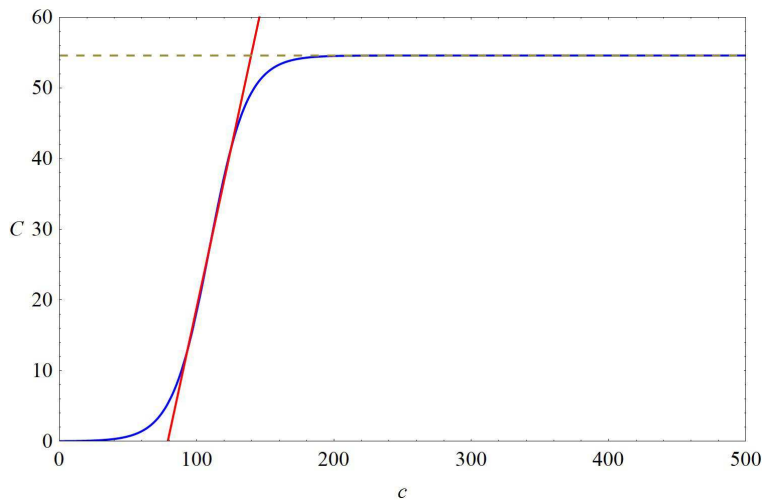}
\caption{The plot is drawn for the value of $s=4$. The horizontal axis is $C$ the former proposals of holographic complexity. The dashed line is the maximum entropy $e^s$. The red line shows the slope of the curve and indicates its linearity in the ramp region.}\label{hp}
\end{figure}

With these parameters, at late times when CV and CA find complexity infinite, $\mathcal{C}$ tends to $e^s$, so it respects the maximum complexity criterion. In addition, where $C\sim s/\omega$ the hyperbolic proposal is reduced to $\mathcal{C}\propto C$, which is linear as Fig. \ref{hp} depicts.

Having shown the linearity, then to fulfill the criteria of holographic complexity we should consider perturbed TFD state. In view of the fact that hyperbolic proposal is linear where $C\sim s/\omega$ , as a consequence it satisfies the perturbed case. However, to show that this proposal can make it in other regions, we expand Eq. (\ref{hp1}) around a generic point $p$ as follows:

\begin{equation} 
\frac{\sinh(\omega C)}{\cosh(\omega C-s)}=\frac{\sinh(\omega p)}{\cosh(\omega p)}+\omega \frac{\cosh(s)}{\cosh^2(\omega p-s)} (C-p)-\omega^2 \frac{\cosh(s)~\tanh(\omega p-s)}{\cosh^2(\omega p-s)} (C-p)^2+... \label{lig}
\end{equation}
Since $\omega=\text{sech}(s)$ is less than one, for $p$ in the ramp region, we have $\mathcal{C}'(p)\gg \mathcal{C}''(p)\gg \mathcal{C}'''(p)\gg ...$ so nonlinear terms are exponentially suppressed and can be discarded. As an example, consider $s=4$, $p$ in the ramp region and $(C-p)\sim \mathcal{O}(1)$, we find the order of terms in right-hand side of \eqref{lig}, respectively to be $\mathcal{C}\sim \mathcal{O}(10)+\mathcal{O}(1)+\mathcal{O}(10^{-3})+\mathcal{O}(10^{-4})+\cdots$. This explains the linearity in the ramp region.

\begin{figure}
\centering
\includegraphics[width=10cm]{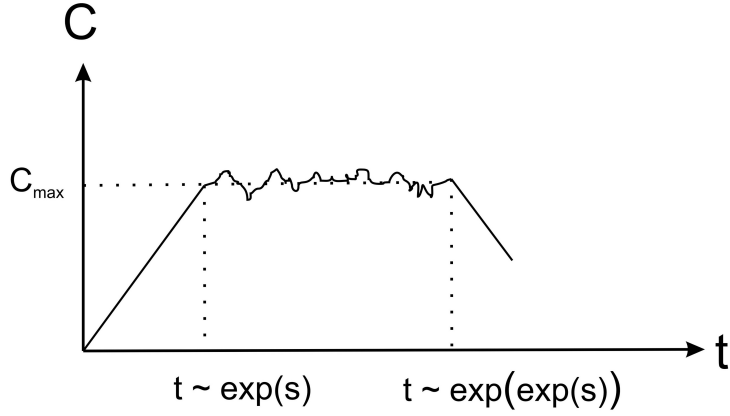}
\caption{The plot of complexity in a quantum circuit goes with complexity ramp and the plateau.}\label{qc}
\end{figure}

There remains one point on the linearity. It is assumed that the linear growth of complexity except at the beginning is continued to its maximum value, which is called the region of complexity ramp, and after reaching the maximum, it fluctuates around the maximum, which is called the plateau and after the time of order double exponential in entropy, it jumps down as Fig. \ref{qc} shows \cite{expo}. While other proposals only represent some parts of the complexity ramp, the hyperbolic proposal can demonstrate the whole ramp region up to the maximum. In the final stage, the complexity jump can not be explained by either proposal.

\subsection{Lloyd bound}\label{lob}
To see how \eqref{hp1} fares better than previous proposals, we find the derivative of Eq. (\ref{np}):
\begin{equation}
\dot{\mathcal{C}}=\frac{d\mathcal{C}}{dt}=\frac{\dot{C}}{\cosh^2(\omega C-s)} \label{hlo}
\end{equation}

Owing to the fact that the maximum value of the term $1/\cosh^2(\omega C-s)$ is one at $C=s/\omega$ and for $C\gg s/\omega$, $\dot{\mathcal{C}}$ rapidly goes away, it follows then:

\begin{equation}
\dot{\mathcal{C}}=\frac{\dot{C}}{\cosh^2(\omega C-s)} \le \dot{C} \label{l1}
\end{equation} 
For heavy black holes, $\omega=\text{sech}(s)$ tends to zero, and as a consequence, $\cosh^2(\omega C-s)$ tends to infinity, which means that Lloyd bound by the hyperbolic proposal at the early stages of complexity growth tends to zero. In other words, if the bound violation in the former proposals occurs at the early stages of complexity growth, then hyperbolic proposal respects the bound. Even if the violation in the former proposals occurs at the middle or late stages of complexity growth, since $\frac{\dot{C}}{\cosh^2(\omega C-s)}< \dot{C}$, hyperbolic proposal behaves better than the former. For lighter black holes $\omega=\text{sech}(s)$ tends to one then at the first stages of complexity growth where $C$ is small, the Lloyd bound in the hyperbolic proposal is similar to the former proposals. However, as complexity grows, $\frac{1}{\cosh^2(\omega C-s)}$ becomes less than one, so we can declare that always $\dot{\mathcal{C}}\le \dot{C}$, which means that hyperbolic proposal in regards to the Lloyd bound, behaves better than former proposals. 

\begin{figure}
\centering
\subfloat[][]{ \includegraphics[width=13cm]{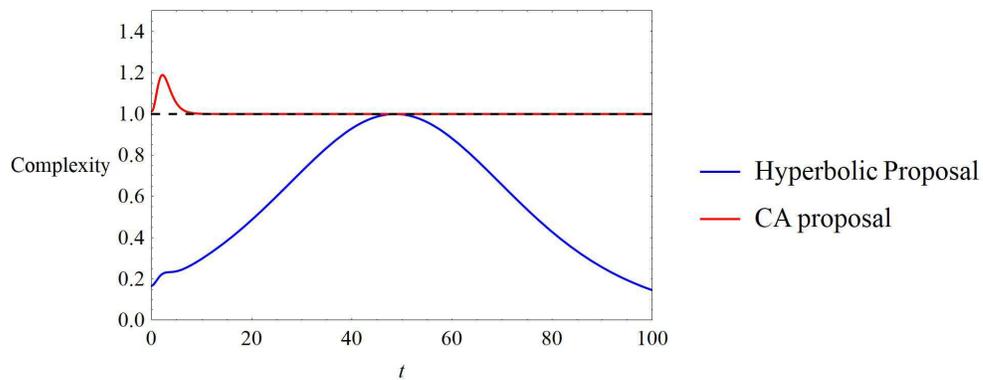}\label{0}}
\vfill
\subfloat[][]{\includegraphics[width=13cm]{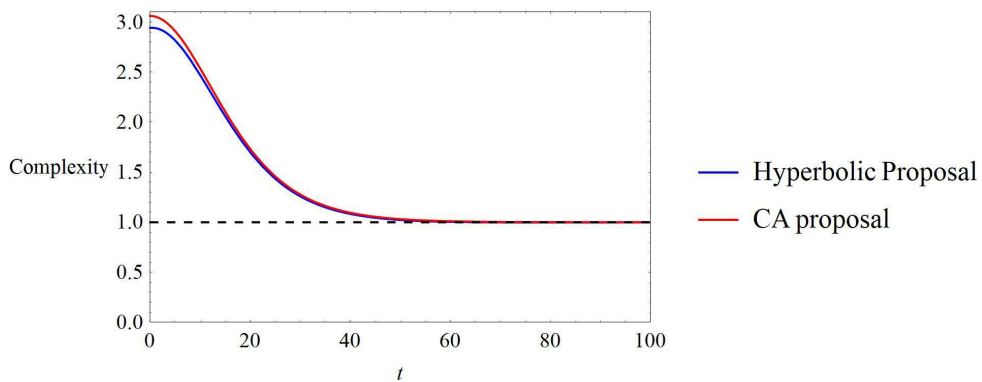}\label{2}}

\caption{Lloyd bound in BTZ black hole by CA and hyperbolic proposals. In panel (a) we set s=1.5 representing a large black hole. Panel (b) with s=0.2 represents a small black hole.}\label{BTZ}
\end{figure}

It is worth comparing the Lloyd bound in AdS-BTZ black hole in CA and the hyperbolic proposal. In \cite{onthe}, Carmi et al. showed that complexity growth rate violates Lloyd bound for AdS-BTZ black holes in the first stages of complexity growth, then we expect that hyperbolic proposal shows the violation in light black holes and respects it in heavy black holes as in Fig. \ref{BTZ}.

\section{Sachdev-Ye-Kitaev Model}\label{syk}
The Sachdev-Ye-Kitaev (SYK) Model is an example of a chaotic quantum system that contains $N$ Majorana fermions $\psi_i$ with a random Hamiltonian \cite{mal}:
\begin{align}
H=\sum_{i_1 < ... < i_a} I_{i_1 ... i_a}\psi_{i_1}... \psi_{i_a}
\end{align}
where the couplings $ I_{i_1 ... i_a}$ are chosen at random from a Gaussian distribution with mean zero and variance $\sigma^2$: 

\begin{align}
\sigma^2=\frac{(a-1)!\mathcal{I}^2}{N^{a-1}}
\end{align}
in which $\mathcal{I}$ is the variance setting parameter. 
In \cite{syk} for $N=2$ fermions, the disorder-averaged complexity when $I$ is chosen to be Gaussian-distributed is defined as:
\begin{equation}
\overline{C}(t)=\int_0^{\infty}dI C_{QM}(t) \frac{e^{-\frac{I^2}{2\sigma^2}}}{\sqrt{2\pi}\sigma}
\end{equation}
where $I=\sqrt{\Sigma_i I_i^2}$ for $I_i$ as couplings of Hamiltonian and $C_{QM}$ is the complexity in the quantum side. For $N=2$ it is given by \cite{syk}:
\begin{equation}
C_{QM}(t)=\frac{\pi}{2}-\frac{4}{\pi}\sum_{n=1}^{\infty}\frac{\cos\big((2n-1)I t\big)}{(2n-1)^2}
\end{equation}

\begin{figure}
	\centering
	\includegraphics[width=10cm]{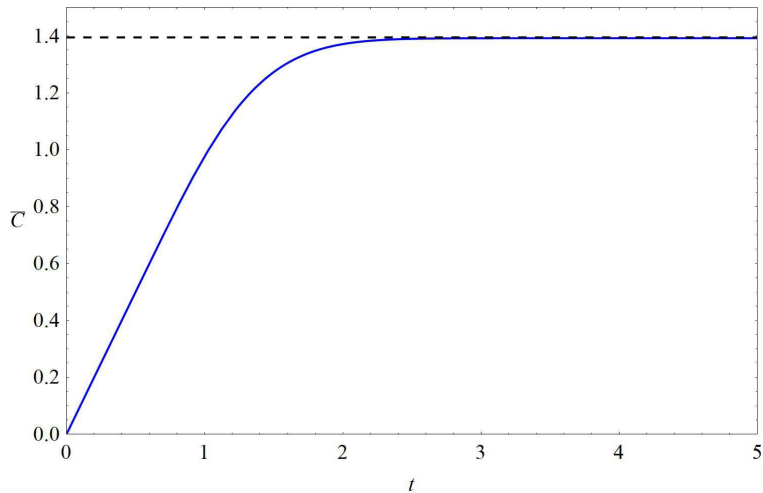}
	\caption{The plot of averaged complexity with $I$ chosen to be Gaussian-distributed.}\label{gauss}
\end{figure}

The holographic dual of the SYK model is JT gravity, which is a two-dimensional quantum gravity with a dilaton \cite{jt}. This gravitational theory is not dual to a quantum theory with a specific Hamiltonian but to an ensemble of theories with various $I$. So it is expected that the holographic complexity may represent the average behavior of the complexity on the quantum side. It is believed that holographic picture breaks down after time scale of exponential in entropy \cite{three lecture,expo}, while in \cite{syk} the averaged complexity is computed from the beginning until times longer than exponential in entropy and is given by:

\begin{equation}
\overline{C}(t)=\frac{2}{\sqrt{\pi}}\sum_{n=1}^{\infty}\frac{1}{(2n-1)^2}\frac{\tanh\big(\frac{(2n-1)^2 t^2 \sigma^2}{4}\big)}{1+\tanh\big(\frac{(2n-1)^2 t^2 \sigma^2}{4}\big)}
\end{equation}
which is plotted in Fig. \ref{gauss}. 

\begin{figure}
	\centering
	\includegraphics[width=10cm]{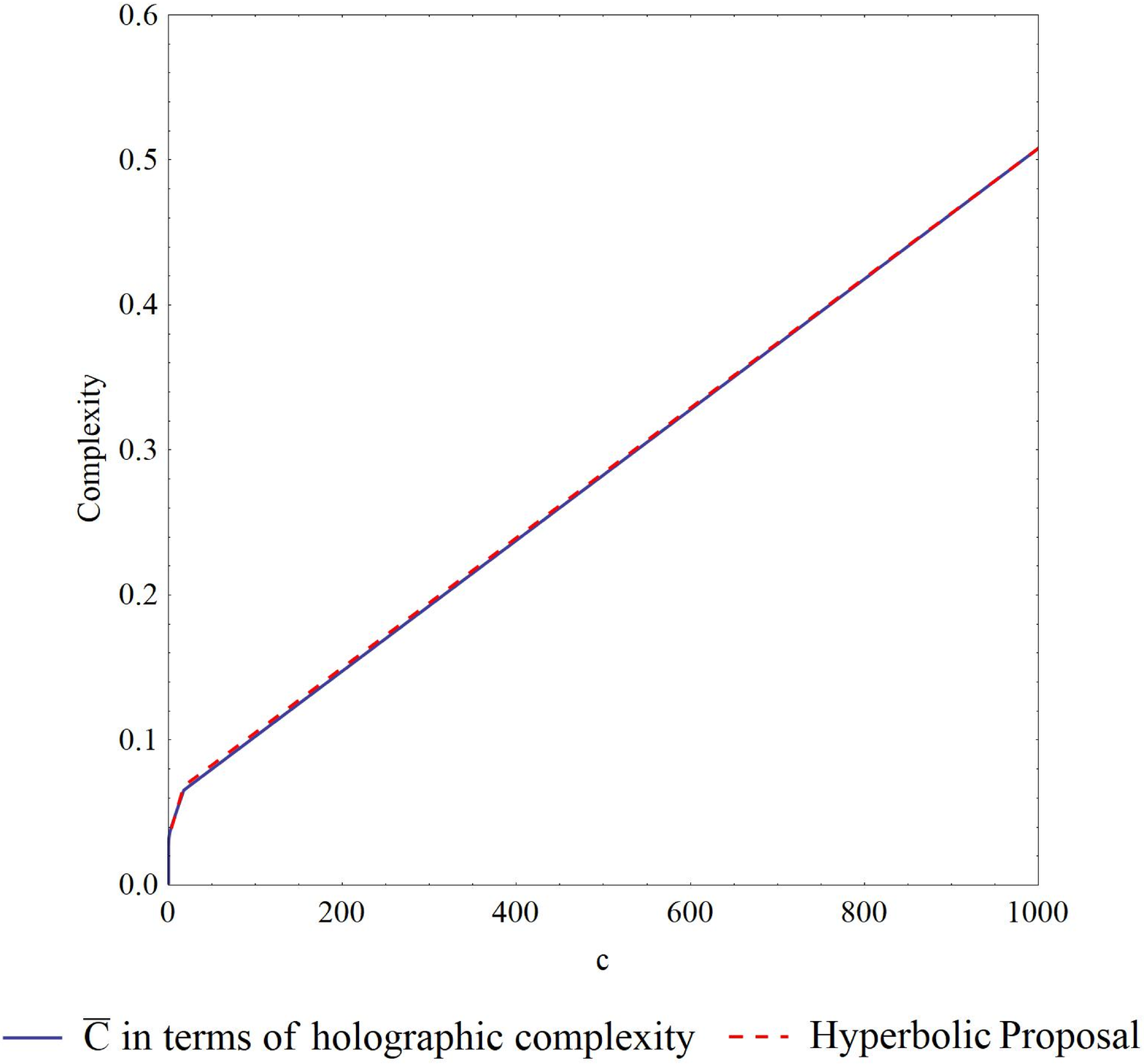}
		\caption{Comparison between hyperbolic proposal and averaged complexity in terms of holographic complexity $\bar{C}\big(C(t)\big)$. The plot is drawn for parameters $\sigma=\gamma=0.009$, $A=0.307$, $\alpha=-0.118$ and $\beta=0.696$ in \eqref{fit} with $\omega=\text{sech}(\beta-\alpha)$. In this vein, the entropy is given by $s=\beta-\alpha$.}\label{comp}
\end{figure}

The plot in Fig. \ref{gauss}, shows that the averaged complexity with $I$ chosen to be Gaussian-distributed has a linear growth and reaches a maximum plateau. This suggests that the averaged complexity may be the very quantity that can represent the right behavior of the complexity in this model \cite{syk}. So one expects that on the bulk side, the holographic complexity may correspond to this averaged quantity. To investigate this, we can find CV holographic complexity for JT gravity (see \cite{jtg}), draw $\bar{C}$ in terms of it and make a comparison between hyperbolic proposal \eqref{hc} and $\bar{C}\big(C(t)\big)$, averaged complexity in terms of holographic complexity. In Fig. \ref{comp}, the solid curve is the averaged complexity and the dotted curve is:
\begin{equation}\label{fit}
	\mathcal{C}=A \frac{\sinh\big(\gamma \omega C -\alpha\big)}{\cosh\big(\gamma \omega C-\beta\big)}
\end{equation}
where we considered $\omega_1=\omega_2=\gamma\omega$ in \eqref{hc} with $\omega=\text{sech}(s)$ and the entropy is $s=\beta-\alpha$. Numerical results show that \eqref{fit} still satisfies the Lloyd bound. 
Notice that we argued that the maximum complexity should be in the order of $\mathcal{O}(e^s)$, we, therefore, expect the numerical factor $A$ to be in the order of  $\mathcal{O}(1)$ or polynomial in $s$. The parameter $\alpha$ reveals that the holographic complexity has a non-vanishing value at the beginning while the averaged complexity vanishes. This behavior could be addressed to the CV and CA proposal at the beginning. It could be seen that either proposal does not represent a linear growth at the beginning while in quantum theory we expect complexity or averaged complexity would be linear, then hyperbolic proposal which uses the former proposals suffers the beginning stages of complexity evolution. Then the remaining nontrivial parameter would be $\omega$ which numerical analysis shows to be corresponding to $\sigma$ in the SYK model. The parameter $\sigma$ is proportional to the slop of the averaged complexity in the complexity ramp \cite{syk} and in holography, the slop of complexity ramp is proportional to $s.T$ where $s$ is entropy and $T$ the temperature, then $\sigma$ is proportional to the temperature.

So it is very suggestive to consider the hyperbolic proposal \eqref{fit} as a holographic dual to the disorder-averaged complexity with $I$ chosen to be Gaussian-distributed, i.e., $\overline{C}=\mathcal{C}$. In this manner, the hyperbolic complexity (\ref{fit}) can go beyond the time scale in CA or CV proposals, and indeed the hyperbolic proposal as a conjecture for averaged complexity can broaden holography beyond that time scale to the complexity plateau region.

\section{Conclusion}
The holographic complexity makes a new avenue to explore black hole physics further. It is expected that any holographic complexity candidate embraces some features: the linear growth, particular behavior under perturbation, the maximum of order exponential in entropy, and the Lloyd bound. The current proposals of holographic complexity do not meet some of these conditions, particularly the maximum value. In other words, they find the maximum value linear in entropy. The reason behind that could be referred to the Poincaré recurrence theorem and the breakdown of the bulk side at late times regime in AdS/CFT. Nevertheless, having the Poincaré recurrences on the bulk side is reachable which makes the expectation that a holographic proposal for complexity would show the maximum value. Moreover, the Lloyd bound in some models is violated. These situations require that the known proposals of holographic complexity should be modified.

We have introduced a new class of holographic complexity conjecture and called it the hyperbolic proposal which is a hyperbolic function of previous proposals in which the maximum complexity of order exponential in entropy is reached. The new proposal is inspired by the behavior of the complexity in BTZ black holes where the complexity could be computed analytically. As a proposal for complexity, the linear growth in time and the behavior under the perturbation have been shown. Although the Lloyd bound in this new proposal is not guaranteed to be satisfied, the complexity growth rate in the hyperbolic proposal is always less than or equal to its former counterparts in CA and CV. It could be seen that in large masses, the complexity growth rate in the hyperbolic proposal is much less than its counterparts, and in light masses, it is comparable to them. Then, if the violation occurs in the early or middle stages of the complexity growth of a large mass regime, the hyperbolic proposal behaves the Lloyd bound much better than the former proposals. 

In SYK models, which are chaotic quantum mechanical systems, the disorder-averaged complexity resembles the hyperbolic proposal. Subsequently, it could be considered that the hyperbolic proposal in such systems is a conjecture for disorder-averaged complexity with $I$ chosen to be Gaussian-distributed, and not the complexity itself. Then it can exhibit beyond the time scale of order exponential in entropy, which has been believed that holographic picture could not go beyond that time scale. Consequently, the hyperbolic proposal as a conjecture for disorder averaged complexity can broaden our holographic perspective to the complexity plateau region.

\section*{Acknowledgement}
We are grateful to Mohsen Alishahiha for useful discussions and reading the manuscript and also to Ahmad Moradpoori for helpful comments.



\end{document}